\documentstyle[emulateapj] {article}

\begin{document}

\title{HST NICMOS Observations of FSC10214+4724}

\author{A. S. Evans\altaffilmark{1,2}, N. Z. Scoville\altaffilmark{1},
N. Dinshaw\altaffilmark{3},
L. Armus\altaffilmark{4},
B. T. Soifer\altaffilmark{1,4}, 
G. Neugebauer\altaffilmark{1}, and
M. Rieke\altaffilmark{5}}

\altaffiltext{1}{Division of Physics, Math, \& Astronomy,
California Institute of Technology, Pasadena, CA 91125}
\altaffiltext{2}{Email Address: ase@astro.caltech.edu}
\altaffiltext{3}{UCO/Lick Observatory, University of California,
Santa Cruz, CA 95064}
\altaffiltext{4}{SIRTF Science Center, California Institute of Technology,
Jet Propulsion Laboratory,
Pasadena, CA 91125}
\altaffiltext{5}{Steward Observatory, University of Arizona, Tucson, AZ 
85721}

\begin{abstract}

High-resolution, 1.10, 2.05, 2.12, and 2.15 $\mu$m imaging of the
gravitationally lensed system FSC10214+4724 are presented. These data
extend Hubble Space Telescope (HST) observations of the lens system to
redder wavelengths, thus providing the highest resolution images to date
of the rest-frame optical and narrow-line (i.e., H$\alpha$+[N II]) regions
of the background quasar.  The length of the arc in the wide-band
continuum images increases with increasing wavelength, and the
H$\alpha$+[N II] emission has a length in between that of the 1.10 and
2.05 $\mu$m.  The structure of the arc changes from having an eastern and
western peak at 1.10 $\mu$m to a single peak in the center at 2.05
$\mu$m.  The changing structure and length of the arc can be understood in
terms of a model where the background quasar consists of a region of
scattered AGN light that dominates at 1.10 $\mu$m (rest-frame 3300\AA),
surrounded by a more extended narrow-line region. An even more extended
red stellar population would thus contribute light at 2.05 $\mu$m
(rest-frame 6200\AA).  In addition, the H$\alpha$+[N II] emission has
structural features similar to the 1.10 $\mu$m emission normalized by the
(predominantly stellar) 2.05 $\mu$m emission, possibly confirming that the
1.10 $\mu$m emission is a superposition of the sources associated with the
line emission (AGN/massive stars) and the red stellar component that
dominates the 2.05 $\mu$m emission.

The counterimage of the lensed quasar is detected in the 1.10 and 2.05
$\mu$m images, and the rest-frame 3300 and 6200\AA$~$ magnifications of the
lensed quasar are calculated to be 50$\pm11$ and 25$\pm6$, respectively,
which translates into rest-frame optical luminosity for the quasar of
$\sim6\times10^9$ L$_\odot$.  These magnification values are lower than
the previously measured magnification of $\sim100$ at rest-frame 2400\AA.
If the dust in the primary lensing galaxy is not affecting the
measurement of the counterimage flux at 2400 and 3300\AA, the
magnification of the quasar appears to decrease with increasing
wavelength.

Flux measurements of the primary lensing galaxy fit the
spectral energy distribution of an unevolving elliptical galaxy at a
redshift of 0.9, consistent with previous determinations of the redshift.

\end{abstract}


\section{Introduction}

Since its discovery (Rowan-Robinson et al. 1991), the high-redshift,
far-infrared selected source FSC10214+4724 has been studied in exhaustive
detail.  Much of the work done prior to 1994 was inspired by the belief
that FSC10214+4724 was in the early stages of formation; large quantities of
dust and star-forming molecular gas were inferred from its high observed
far-infrared (Rowan-Robinson et al.  1991) and CO luminosities (Brown \&
Vanden Bout 1992; Solomon, Downes, \& Radford 1992). Evidence for a buried
active galactic nucleus (AGN) was also deduced from imaging polarimetry
(Lawrence et al 1993) and rest-frame optical emission-line diagnostics
(Elston et al. 1994; Soifer et al. 1995; Iwamuro et al. 1995), indicating
that, in addition to starlight, much of the observed luminosity ($L_{\rm ir}
[8-1000\micron] \sim 10^{14}$ L$_\odot$) might be reprocessed
AGN light.  Growing suspicion that FSC10214+4724 might be gravitationally
lensed by foreground galaxies (Elston et al.  1994; Matthews et al. 1994;
Trentham 1995) led to a series of observations employing various
techniques to achieve the highest possible resolution (e.g., Graham \& Liu
1995; Broadhurst \& Leh\'{a}r 1995; Serjeant et al.  1995; Close et al.
1995).  However, it was not until FSC10214+4724 was observed with the HST
Wide-Field Planetary Camera 2 (WFPC2) that its nature was conclusively 
resolved; the 0.8 $\mu$m (rest-frame 2400\AA) image showed
unambiguously the lensed quasar, counterimage, and lensing galaxies
(Eisenhardt et al.  1996, hereafter E96).

The fact that FSC10214+4724 is a gravitationally lensed system makes it no
less intriguing a source. Indeed, lensing has made it possible to study
the properties of distant galaxies that would otherwise be too faint to
observe. Further, because the likelihood of lensing increases with the
increased distance (and thus the increased volume of intervening galaxies)
of the background galaxy to the observer, many galaxies found at the
highest redshifts may be lensed as well.

The availability of the recently installed Near-Infrared Camera and
Multiobject Spectrometer (NICMOS) on HST has made it possible to obtain
high resolution ($\sim 0.1\arcsec-0.2\arcsec$) images of the continuum and
narrow emission-line regions in the lensed quasar at rest-frame
wavelengths redward of 2700\AA.  Specifically, optical radiation emitted
from a source at $z\sim2.3$ is redshifted to near-infrared wavelengths in
the present epoch, and thus a program to image the lensed quasar at
rest-frame optical wavelengths and in H$\alpha$+[N II] with NICMOS has
been carried out.  Because of the effects of dust and the possible
differences in the size scales of the continuum and line emission regions,
the extent and structure of the arc will likely change as a function of
wavelength. Of equal importance is the amount to which the intrinsic
luminosity of the quasar is amplified as a function of wavelength.  This
effect is due to the fact that different regions of the background object
have different colors, and are thus amplified by different amounts. For
FSC10214+4724, the magnification is determined from the fluxes of the arc
and counterimage (E96).  Finally, these observations provide additional
photometric data to complement WFPC2 and ground-based measurements of the
lensed quasar and lensing galaxies.

The paper is divided into five sections. The HST observations and data
reduction are summarized in \S2. A detailed description of arc is provided
in \S3, along with a brief description of the photometry. In \S4, the
properties of the lensed quasar and the lensing galaxies are deduced.
Section 5 summarizes the paper.

Throughout this paper, $H_0 = 75$ km s$^{-1}$ Mpc$^{-1}$ and a $q_0 = 0.5$
are adopted, such that for a source at a $z=0.9$, 5.6 kpc subtends
$1\arcsec$ on the sky.  For sources at redshifts of 0.9 and 2.286, the
luminosities distances are 4170 and 11780 Mpc, respectively.

\section{Observations and Data Reduction}

HST observations of FSC10214+4724 were obtained in a single orbit on 1997
October 27 (UT) using camera 2 of NICMOS.  Camera 2 consists of a
256$\times$256 HgCdTe array with pixel scales of 0.0762$\arcsec$ and
0.0755$\arcsec$ per pixel in $x$ and $y$, respectively, providing a
$\sim19.5\arcsec\times19.3\arcsec$ field of view (Thompson et al. 1998).
Images were obtained using the wide-band ($\Delta\lambda_{\rm FWHM} \sim
0.6$ $\mu$m) F110W (1.10 $\mu$m) and F205W (2.05 $\mu$m) filters,
providing a full width at half the maximum flux (FWHM) for a point
source of 0.11$\arcsec$ and 0.20$\arcsec$, respectively.  Observations
were done by executing a 4-point spiral dither per filter setting; the
step size used was 25.5 pixels (1.91$\arcsec$).  At each dither position,
non-destructive reads (MULTIACCUM) were obtained, with integration times
of 96 seconds per dither position.  The total integration time per
wide-band filter setting was thus 384 seconds.

Two narrow-band ($\Delta\lambda_{\rm FWHM} \sim 0.02$ $\mu$m) images of
FSC10214+4724 were also taken. The F215N filter (2.15 $\mu$m) is centered
at the wavelength of the redshifted H$\alpha$+[N II] emission, and the
F212N filter (2.12 $\mu$m) is centered at 6450\AA$~$ in the rest-frame of
the lensed quasar (i.e., continuum only).  The FWHM of a point source in
both filters is $\sim0.21\arcsec$.  Observations with the narrow-band
filters were done in the same fashion as the wide-band filters, with
longer integration times of 120 seconds at each dither position. The total
integration time per filter setting was thus 480 seconds.  Finally, dark
exposures were taken using the same MULTIACCUM sequences executed for the
quasar observations.

Reduction of the data was done with IRAF.  The dark was first created,
then the NICMOS data were dark subtracted, flatfielded and corrected for
cosmic rays using the IRAF pipeline reduction routine CALNICA (Bushouse
1997).  The dithered images were then shifted and averaged using the
DRIZZLE routine in IRAF (e.g. Hook \& Fruchter 1997). The plate scales of
the final ``drizzled'' images are 0.0381$\arcsec$ and 0.0378$\arcsec$ per
pixel in $x$ and $y$, respectively. The resultant images are shown in
Figure 1. 

Of key interest to this project is the relative positioning of the arc
and the counterimage to the lensing galaxies as a function of wavelength.
To check the accuracy of the pointing during the observations, the
relative positions of Source 2 (see \S 3.1) were measured in the wide-band
images using the IRAF routine IMEXAMINE; the centroid of Source 2
in the two images differed by only 0.04 ($0.0015\arcsec$) of a pixel.

Finally, flux calibration of the images were done using the scaling
factors 2.28$\times10^{-6}$, 1.55$\times10^{-6}$, 4.07$\times10^{-5}$, and
4.48$\times10^{-5}$~Jy (ADU/sec)$^{-1}$ at 1.10, 2.05, 2.12, and 2.15
$\mu$m, respectively (Rieke et al. 1998). The corresponding magnitudes
were calculated using the zero-points 1909, 707, 686 and 680 Jy (Rieke et
al. 1998).

\section{Results}

For ease in direct comparison, the sources within the field of view
(Figure 1) have been numbered in the same manner as done by Matthews et
al.  (1994) and E96.\footnote{References to Sources 1-4 should be made
in accordance with the {\it IAU} naming convention FSC 10214+4724:M (number).
Likewise, Source 5 should be referred to as FSC10214+4724:E 5.}  All of the
sources previously observed by E96 were detected in the 1.10 and 2.05
$\mu$m images.  Source 1 is observed to be an arc with a FWHM of
$\sim0.6\arcsec$ in length and unresolved in width.  A more detailed
description of the arc will be provided in \S3.1.  The primary lensing
galaxy (Source 2) has a compact nuclear region and an underlying, low
surface brightness envelope; the FWHM of Source 2 is $\sim0.29\arcsec$ in
the 2.05 $\mu$m image.  Two additional galaxies are visible; the secondary
lensing galaxy (Source 3: see E96) is an asymmetric galaxy with a FWHM of
$\sim0.35\arcsec$ in the 2.05 $\mu$m image and an unresolved nucleus at
its western end.  Source 4, which is a relatively faint, compact (FWHM
$\lesssim0.20\arcsec$) galaxy in the 2.05 $\mu$m image, appears to be a
highly inclined galaxy in the WFPC2 image (E96).  Source 5, the
counterimage of the lensed quasar, appears as a faint northern extension
of Source 2, and will be discussed in more detail in \S 4.1.

\subsection{Arc Structure}

Figure 2 consists of a contour plot of the 0.8 $\mu$m image of Sources 1
and 2 taken from E96, and contour plots of the 1.10, 2.05, and 2.15 $\mu$m
images.  The tangential extent of the arc differs in the wide-band images;
the full width at 10\% the maximum flux density of Source 1 subtends an
angle of $\sim46^{\rm o}$ at 0.8 $\mu$m relative to the position of the
primary lensing galaxy (Source 2), $\sim46^{\rm o}$ at 1.10 $\mu$m, and
$\sim71^{\rm o}$ at 2.05 $\mu$m. Given that Source 2 is $\sim 1.2\arcsec$
north of Source 1 (as measured from the peak of Source 1 at 2.05 $\mu$m),
the arc has observed lengths of $\sim0.95\arcsec$, $0.95\arcsec$, and
$1.5\arcsec$ at 0.8, 1.10, and 2.05 $\mu$m, respectively.

Using the same criteria as above, the narrow-band, 2.15 $\mu$m image of
the arc subtends an angle of $\sim61^{\rm o}$ relative to the primary
lensing galaxy, and is thus $1.3\arcsec$ in length.  As expected, the arc
is unresolved in width at 2.15 $\mu$m.  In the 2.12 $\mu$m continuum
image, the arc is relatively faint, indicating that continuum emission
comprises only a small fraction of the 2.15 $\mu$m flux.  Further, aside
from the lensed quasar, there appear to be no strong ($m_{2.15} < 18.3$
mag) emission-line sources at the redshift of Source 1 present in the
field.

Figure 3 shows close-up, contour plots of Source 1.  As noted by E96, the
0.8 $\mu$m image of Source 1 is asymmetric, having a primary eastern peak
separated from a secondary western peak by $\sim0.24\arcsec$ (Figure 3a).
The 1.10 $\mu$m (i.e., rest-frame 3300\AA) emission from the arc has an
asymmetric appearance similar to the 0.8 $\mu$m (rest-frame 2400\AA$~$)
emission; the arc has an asymmetric appearance, with a major peak (P1) on
the eastern end of the arc separated by $\sim0.27\arcsec$ from a faint,
minor peak (P2) on the western end.  In contrast to the 0.8 and 1.10
$\mu$m emission, the 2.05 $\mu$m (rest-frame 6200\AA) emission has a
nearly symmetric appearance, with a single peak at the center of the arc.
The position of the primary 1.10 $\mu$m and the 2.05 $\mu$m peaks differ
by 0.10$\arcsec$.

Figure 3f shows the ratio of the 1.10 and 2.05 $\mu$m images.  The image
consists of a two-component arc with fainter emission bridging the
components. The primary peak of the image is marginally shifted
(0.04\arcsec) eastward of the primary 1.10 $\mu$m peak (P1), and the
secondary peak is shifted a similar amount from the secondary 1.10 $\mu$m
peak (P2).

Figure 3g and 3h show the contour plots of the 2.15 $\mu$m image and the
difference of the 2.15 and 2.12 $\mu$m images (i.e., H$\alpha$+[N II]).
Because the peak of the 2.12 $\mu$m image appears shifted
$\sim0.10\arcsec$ west of the 2.05 $\mu$m peak, an additional check
of the structure of the H$\alpha$+[N II] emission was done by scaling the
2.05 $\mu$m image to the flux of the 2.12 $\mu$m image, then subtracting
it from the 2.15 $\mu$m image; the resultant arc showed only marginal
changes from the subtraction using the 2.12 $\mu$m image.  The
similarities between the images shown in Figure 2g and 2h are due to the
large contribution of line emission to the overall flux density and
structure of the 2.15 $\mu$m emission.

\subsection{Photometry}

Table 1 lists the magnitudes derived from the images in Figure 1, as well
as magnitudes in the wavelength range 0.7--2.2 $\mu$m compiled from the
literature. The magnitudes for all of the sources are consistent with
previous ground-based measurements.  Both of the wide-band NICMOS images
of Source 1 are composed of continuum and line emission; strong Ne V and
Ne III emission have been detected in the wavelength range 1.1--1.3 $\mu$m
(Soifer et al.  1995; Iwamuro et al. 1995), and H$\alpha$+[N II] emission
has been observed at 2.15 $\mu$m (Elston et al.  1994; Soifer et al.
1995). The percentage of line contribution to the 2.05 $\mu$m flux of
Source 1 can be calculated from the narrow-band images. Subtracting the
2.12 $\mu$m continuum image from the 2.15 $\mu$m and measuring the flux of
the resultant emission-line image, the H$\alpha$+[N II] flux is calculated
to be 4.3$(\pm0.4)\times10^{-18}$ W m$^{-2}$, 30--50\% lower than the
values of 6$\times10^{-18}$ W m$^{-2}$ and 7$\times10^{-18}$ W m$^{-2}$
determined by Matthews et al. (1994) and Elston et al. (1994),
respectively.  Thus, H$\alpha$+[N II] comprises $\sim$12\% of the 2.05
$\mu$m flux of Source 1. This percentage is consistent with the
approximate value of 13\% calculated from the near-infrared spectrum of
FSC10214+4724 by Soifer et al. (1995). Similarly, using the H$\alpha$+[N
II] flux in combination with the Ne V and Ne III to H$\alpha$+[N II] flux
ratio of 1.1 determined by Soifer et al. (1995), the Neon emission lines
are calculated to comprise $\sim$8\% of the 1.10 $\mu$m flux.

\section{Discussion}

Both the length and structure of the continuum and line emission of Source
1 can be explained in terms of the relative sizes of the emission regions,
the structure of the emission regions, and their location near the cusp of
a caustic (i.e., line of infinite magnification:  see Blandford \& Narayan
1992). Given that Source 1 is a dust enshrouded quasar (\S 1), it is very
likely that a substantial fraction of the luminosity from Source 1,
especially at bluer wavelengths, is scattered/reprocessed AGN light. The
length of the wide-band continuum emission has been shown to increase as a
function of wavelength (\S 3.1), indicating that the light emitted at
longer wavelengths is closer to the caustic than the shorter wavelength
light.  Physically, this can be understood if the 0.8--1.10 $\mu$m light
emanates predominantly from regions of scattered AGN light, and the 2.05
$\mu$m light emanates from the underlying, red stellar population of
Source 1 which is more extended than the scattered light region and has a
substantial cross-section on or near the caustic.  By comparison, the
H$\alpha$+[N II] emission, which traces light from the narrow-line regions
(e.g. Osterbrock 1989), has a length in between that of the 0.8--1.10
$\mu$m and 2.05 $\mu$m emission, indicating that the narrow-line region is
more extended than the scattered AGN light region, but not as extended as
the stellar region traced by the 2.05 $\mu$m emission.

The change in the structure along the arc is indicative of variations in
the morphology of Source 1 as a function of wavelength.  While the 0.8 and
1.10 $\mu$m emission have an eastern and western peak, the fact that the
2.05 $\mu$m emission has only one peak at the center of the arc may be a
result of the red stellar emission being more extended than or displaced
relative to the 0.8--1.10 $\mu$m emission.  Further, the two-component
morphology of the emission-lines is similar to the ratio of the 1.10 and
2.05 $\mu$m emission.  Normalizing the 1.10 $\mu$m emission by the 2.05
$\mu$m emission removes structure at 1.10 $\mu$m caused by the red stellar
population.  Thus, the similarities between the structure of the line
emission and the 1.10 $\mu$m / 2.05 $\mu$m ratio may confirm that the 1.10
$\mu$m emission is a superposition of a blue component associated with the
emission-line emission and a red stellar component that dominates at 2.05
$\mu$m. Such superpositions have also been modeled in radio galaxies at
$z\sim1$, where the images of the galaxies at bluer wavelength appear to
be comprised of an elongated component, as well as a symmetric component
similar in shape to the symmetric images of the galaxies at redder
wavelengths (Rigler et al. 1992).

\subsection{Magnification}

Figure 4 shows 0.8, 1.10, and 2.05 $\mu$m contour plots of Source 2 and
the counterimage (Source 5). There appears to be a marginal shift in the
centroid of Source 5 at 2.05 $\mu$m relative to 0.8 and 1.10 $\mu$m; such
a shift may simply be an artifact of low signal-to-noise, or it may be
further evidence that the morphology of the 0.8--1.10 and 2.05 $\mu$m
emission from the quasar are different, thus causing the counterimage to
appear at a slightly different location on the image plane.

In order to determine the 1.10 and 2.05 $\mu$m fluxes of the counterimage,
which is necessary for determining the magnification of the quasar, the
flux in a 0.38$\arcsec$ diameter aperture centered on the counterimage was
measured, then fluxes were measured in seven apertures positioned the same
distance from the center of Source 2 as the counterimage aperture.  The
rms of the eight positions was then used to determine the rms of the
measured flux of Source 5.  The flux densities were determined to be
0.76$\pm0.11$ and 3.3$\pm0.78$ $\mu$Jy at 1.10 and 2.05 $\mu$m,
respectively.  The total magnification of FSC10214+4724 is simply the
ratio of the arc to counterimage flux densities (E96), thus the
magnification of the lensed quasar is 50$\pm11$ and 25$\pm6$ at rest-frame
3300 and 6200\AA, respectively. If the assumption is made that the
emission emanates from a uniformly illuminated source (Figure 5 of E96),
the source of the emission is $\sim$100 pc (0.015$\arcsec$) in radius at
3300\AA, and $\sim$300 pc (0.04$\arcsec$) in radius at 6200\AA. Further,
correcting the observed luminosities of the quasar for the lensing factors
yields rest-frame 3300 and 6200\AA$~$ luminosities of
$2.8(\pm0.6)\times10^9$ and $6.3(\pm1.7)\times10^9$ L$_\odot$. Typical
rest-frame optical luminosities of ultraluminous infrared galaxies are
$\sim1\times10^{10}$ L$_\odot$, but variations in their rest-frame optical
luminosities result, in part, from dust obscuration.

The magnification of the lensed quasar has previously been determined at
wavelength blueward of rest-frame 3000\AA.  E96 calculated a rest-frame
2400\AA$~$ magnification of $\sim100$, and Nguyen et al. (1998) has
computed a lower limit of the rest-frame 1300\AA$~$ magnification of
$\sim250$. Thus, the magnification of the lensed quasar appears to be
decreasing with increasing wavelength. However, as Nguyen et al.  (1998)
point out, while the arc is most likely too far from the primary lensing
galaxy (Source 2) for its measured flux to be diminished by dust in the
lensing galaxy, the proximity of the counterimage to the nucleus of the
Source 2 may mean that the flux of the 1300-3300\AA$~$ counterimage, and
thus the flux ratio of the arc to counterimage at 1300-3300\AA, are
heavily affected by dust in the primary lensing galaxy.  By the same
argument, the radiation from the lensed quasar which is detected in the
2.05 $\mu$m filter has a wavelength of 1.1 $\mu$m as it passes near the
primary lensing galaxy, and thus is unaffected by dust.

\subsection{The Redshift of the Lensing Galaxies}

As mentioned in \S3.2, the measured fluxes of all of the sources agree
with previous ground-based measurements.  Of particular interest is the
redshift and Hubble type of the lensing galaxies as derived from their
fluxes and the morphologies. The NICMOS data presented here are consistent
with the assertion that Source 2 is an early-type galaxy at a redshift of
0.9 (see discussion of the spectral energy distribution of Source 2 in
Appendix A of E96).  While such a straightforward interpretation of the
spectral energy distribution (SED) of Source 3 is not possible, the close
proximity of the two galaxies and the similarity in their observed extent
indicate that Source 3 is very likely a companion of Source 2. If both
galaxies are at the same redshift, they have a projected separation of
$\sim14$ kpc and size scales (FWHM) of $\sim1.8$ kpc, and it is very
likely that the asymmetric appearance of Source 3 results from a tidally
interaction with Source 2.  The optical luminosities of Source 2 and 3, as
derived from the observed 1.10 $\mu$m magnitudes, are $4.5\times10^9$ and
$2.3\times10^9$ L$_\odot$, respectively, which is comparable to the
luminosities of the bulges of local spiral galaxies (M31 and the Milky Way
galaxies) and of low luminosity elliptical galaxies, but an order of
magnitude lower than the average luminosity of present day elliptical
galaxies (i.e., Nieto et al.  1990).

\section{Summary}

High-resolution near-infrared imaging of the lens system FSC10214+4724 has
been presented.  The observations have provided the highest resolution
images to date of the rest-frame optical and narrow-line region emission
from the lensed quasar.  The following conclusions are reached:

\noindent
(1) The length of the wide-band continuum emission (Source 1) increases
with increasing wavelength.  The full width at 10\% the maximum flux level
of Source 1 is 0.95$\arcsec$ in length at 1.10 $\mu$m and 1.5$\arcsec$ in
length at 2.05 $\mu$m. In comparison, the length of the 0.8 $\mu$m arc
(E96) is also 0.95$\arcsec$.  Thus, if the 0.8--1.10 $\mu$m emission
occurs mostly from a region of scattered AGN light, the 2.05 $\mu$m
emission may be dominated by red stellar light from a more extended region
having a cross-section that overlaps or is near the caustic.

\noindent
(2) The 1.10 $\mu$m image of the arc has a primary eastern and secondary
western peak, similar to the 0.8 $\mu$m emission. In contrast, the 2.05
$\mu$m emission is symmetric, having a peak at the center of the arc.
This indicates that the red stellar emission is displaced from the
emission from scattered AGN light.

\noindent
(3) The  H$\alpha$+[N II] emission has a length in between that of the 1.1
and 2.05 $\mu$m emission.  This may indicate that the narrow-line region
is more extended than the scattered AGN light region, but not as extended
as the red stellar distribution.  Further, the H$\alpha$+[N II]
emission-line image of the arc appears to have a structure similar to the
1.10 $\mu$m emission normalized by the (predominantly stellar) 2.05 $\mu$m
emission, consistent with the idea that the 1.10 $\mu$m emission is a
superposition of sources associated with the emission lines and the
stellar component that dominates the 2.05 $\mu$m emission.

\noindent
(4) The H$\alpha$+[N II] emission in Source 1 has an observed flux of
4.3$(\pm0.4)\times10^{-18}$ W m$^{-2}$. This line emission is calculated
to comprise 12\% of the wide-band 2.05 $\mu$m flux of the lensed quasar.
An 8\% level of emission-line contamination is deduced for the 1.10 $\mu$m
flux.

\noindent (5) The rest-frame 3300 and 6200\AA$~$ magnifications of the
lensed quasar are estimated to be 50$\pm11$ and 25$\pm6$, respectively.
Thus, the quasar is determined to have a rest-frame optical luminosity of
$\sim6\times10^9$ L$_\odot$.

\noindent
(6) The measured flux densities of the primary lensing galaxy (Source 2)
are consistent with previous near-infrared measurements and support the
idea that it is an early-type galaxy at a redshift of 0.9.

\acknowledgements
ASE thanks C. Fassnacht, J. Carpenter, J. Surace, and B. Stobie for
useful discussion and assistance. We also thank the referee for many
useful comments.  This research was supported by NASA grant NAG
5-3042, and the observations were obtained with the NASA/ESA Hubble
Space Telescope operated by the Space Telescope Science Institute
managed by the Association of Universities for Research in Astronomy
Inc. under NASA contract NAS5-26555.

\begin{deluxetable}{crrrrrrrrr}
\pagestyle{empty}
\tablewidth{0pt}
\tablecaption{Magnitudes for Sources in FSC10214+4724 Field}
\tablehead{
\multicolumn{1}{c}{Source} &
\multicolumn{1}{c}{$m_{0.70}$\tablenotemark{a}} &
\multicolumn{1}{c}{$m_{0.79}$\tablenotemark{b}} &
\multicolumn{1}{c}{$m_{1.10}$\tablenotemark{c}} &
\multicolumn{1}{c}{$m_{1.25}$\tablenotemark{d}} &
\multicolumn{1}{c}{$m_{1.6}$\tablenotemark{d}} &
\multicolumn{1}{c}{$m_{2.05}$\tablenotemark{c}} &
\multicolumn{1}{c}{$m_{2.17}$\tablenotemark{d}} &
\multicolumn{1}{c}{$m_{2.12}$\tablenotemark{c}} &
\multicolumn{1}{c}{$m_{2.15}$\tablenotemark{c}}\nl}
\startdata
1 & 20.72$\pm$0.02\tablenotemark{e} & 20.44\tablenotemark{f} &
19.2$\pm$0.1\tablenotemark{g} & 19.0\tablenotemark{g} & 16.9$\pm$0.02 &
17.3$\pm$0.1\tablenotemark{h} & 17.4\tablenotemark{h} & 17.2$\pm$0.2 &
15.4$\pm$0.1\tablenotemark{h} \nl

2 & 22.93$\pm$0.12 & 20.3 & 20.1$\pm$0.1 & 19.4 & 18.51$\pm$0.06 &
18.0$\pm$0.1 & 17.6 & \nodata & \nodata \nl

3 & 23.13$\pm$0.25 & 22.98 & 21.6$\pm0.2$\tablenotemark{e} & 20.7 &
19.52$\pm$0.16 & 18.9$\pm$0.1 & 18.5 & \nodata & \nodata \nl

4 & \nodata        & 23.58 & 22.6$\pm0.2$\tablenotemark{e,i} & 22.4 &
\nodata & 20.0$\pm$0.1\tablenotemark{i} & 20.0 &\nodata  & \nodata \nl

5 & \nodata & 25.5 & 23.5$\pm$0.2\tablenotemark{j} & \nodata & \nodata &
20.8$\pm$0.2\tablenotemark{j} & \nodata

& \nodata & \nodata \nl 

\enddata
\tablenotetext{a}{From Elston et al. (1994).} 

\tablenotetext{b}{From Eisenhardt et al. (1996).}

\tablenotetext{c}{This paper. Unless otherwise noted, all 1.10, 2.05,
2.12, and 2.15$\mu$m magnitudes have been calculated using a 1.37$\arcsec$
diameter aperture.}

\tablenotetext{d}{From Matthews et al. (1994).}

\tablenotetext{e}{Contains emission from CIII] $\lambda$1909 and Ne IV]
$\lambda$2424 lines.}

\tablenotetext{f}{Contains emission from Ne IV] $\lambda$2424 lines.}

\tablenotetext{g}{Contains emission from [Ne V] $\lambda\lambda$ 3346,3426
and [Ne III] $\lambda\lambda$ 3869,3967 lines.}

\tablenotetext{h}{Contains emission from H$\alpha$+[N II]
$\lambda\lambda$6548,6583 lines.}

\tablenotetext{i}{Calculated using a 0.68$\arcsec$-diameter aperture.}

\tablenotetext{j}{Calculated using a 0.38$\arcsec$-diameter aperture.}

\end{deluxetable}

\vfill\eject
\centerline{Figure Captions}

\vskip 0.06in

\noindent
Figure 1a--d: NICMOS 1.10, 2.05, 2.12, and 2.15 $\mu$m images of the field
of FSC10214+4724. The arc (Source 1) is unresolved in width in all the
images.  Note that the noise in the top left-hand corner of each image is
due to the coronagraphic hole on camera 2 of NICMOS.

\vskip 0.05in

\noindent
Figure 2. Contours of Sources 1 and 2. (a) The wide-band 0.8 $\mu$m image
taken from E96. The peak flux density is 0.22 $\mu$Jy. (b,c) The wide-band
1.10 and 2.05 $\mu$m images, with peak flux densities of 0.27 and 0.5
$\mu$Jy.  (d) The narrow-band image, with a peak flux density of 2.4
$\mu$Jy. All contours are displayed at 10, 21, 32, 43, 54, 66, 77, 88, and
99\% of the peak image flux.

\vskip 0.05in
 
\noindent
Figure 3. (a,b) Wide-band 0.8 and 1.10 $\mu$m contours of Source 1, with
peak fluxes densities of 0.37 and 0.48 $\mu$Jy. (c-h) Contours of Source 1
smoothed to a resolution of $\sim0.23\arcsec$.  (c,d,e) The wide-band 0.8,
1.10 $\mu$m and 2.05 $\mu$m images, with peak flux densities of 0.22, 0.27
and 0.50 $\mu$Jy.  (f) The ratio of the 1.10 and 2.05 $\mu$m images, with
peak flux density ratio of 0.52.  (g) The narrow-band 2.15 $\mu$m image,
with a peak flux density of 2.4 $\mu$Jy.  (h) The continuum-subtracted
H$\alpha$+[N II] image, with a peak flux density of 2.1 $\mu$Jy. All
contours are displayed at 60, 65, 70, 75, 80, 84, 89, 94, and 99\% of the
peak image flux.

\vskip 0.05in

\noindent
Figure 4. Contour plots of Sources 1, 2, and 5 at 0.8, 1.10, and 2.05
$\mu$m.  The images have been gaussian smoothed to a resolution of
0.23$\arcsec$, then boxcar smoothed 3$\times$3 (0.8 $\mu$m image) and
4$\times$4 (1.10 and 2.05 $\mu$m images) pixels. The contour levels in
each image have been chosen to highlight Source 5, and thus correspond to
nine linearly spaced contour levels over the flux densities ranges
0.0090--0.014, 0.011--0.020, and 0.038--0.085 $\mu$Jy for 0.8, 1.10, and
2.05 $\mu$m , respectively.

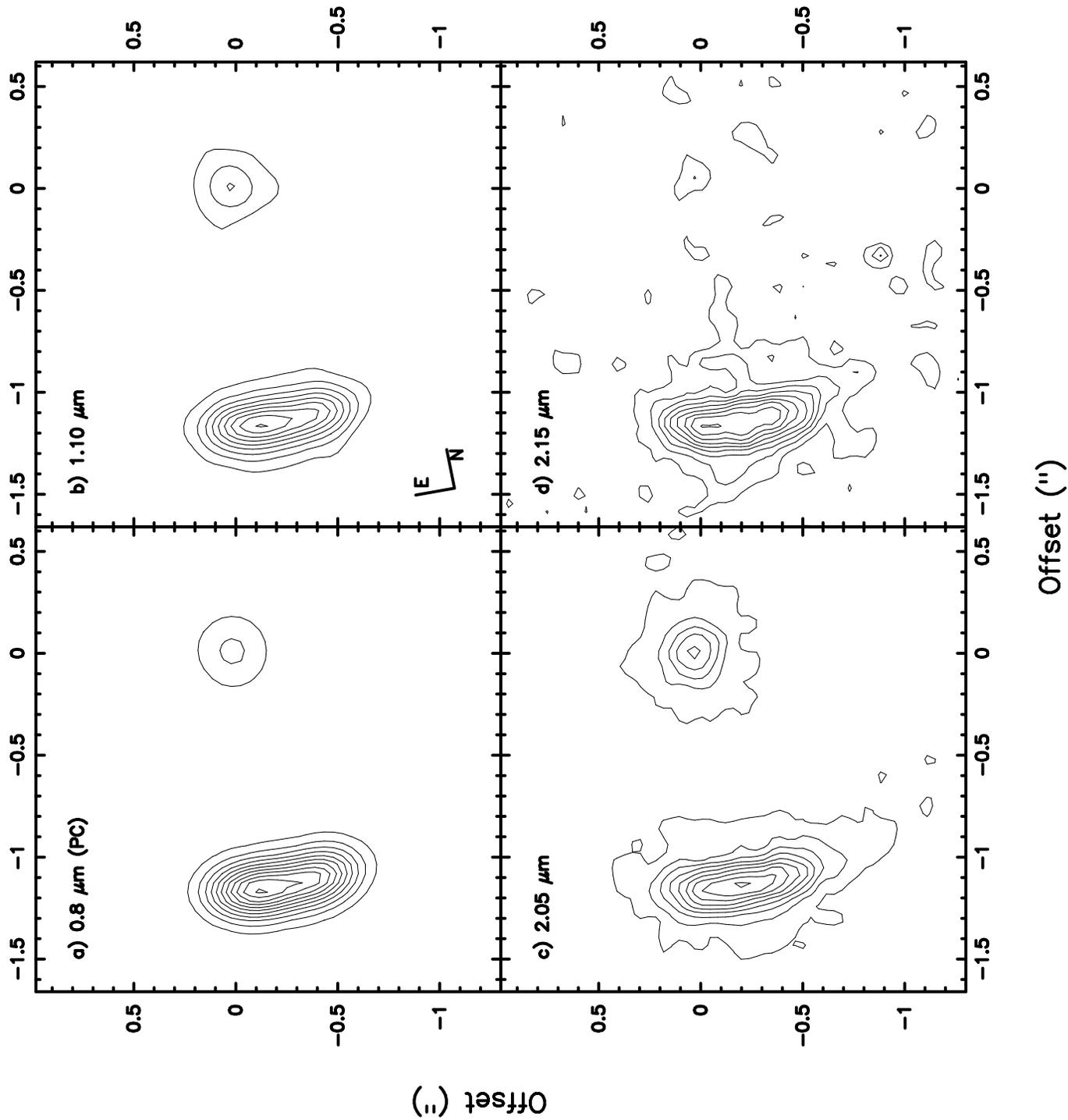
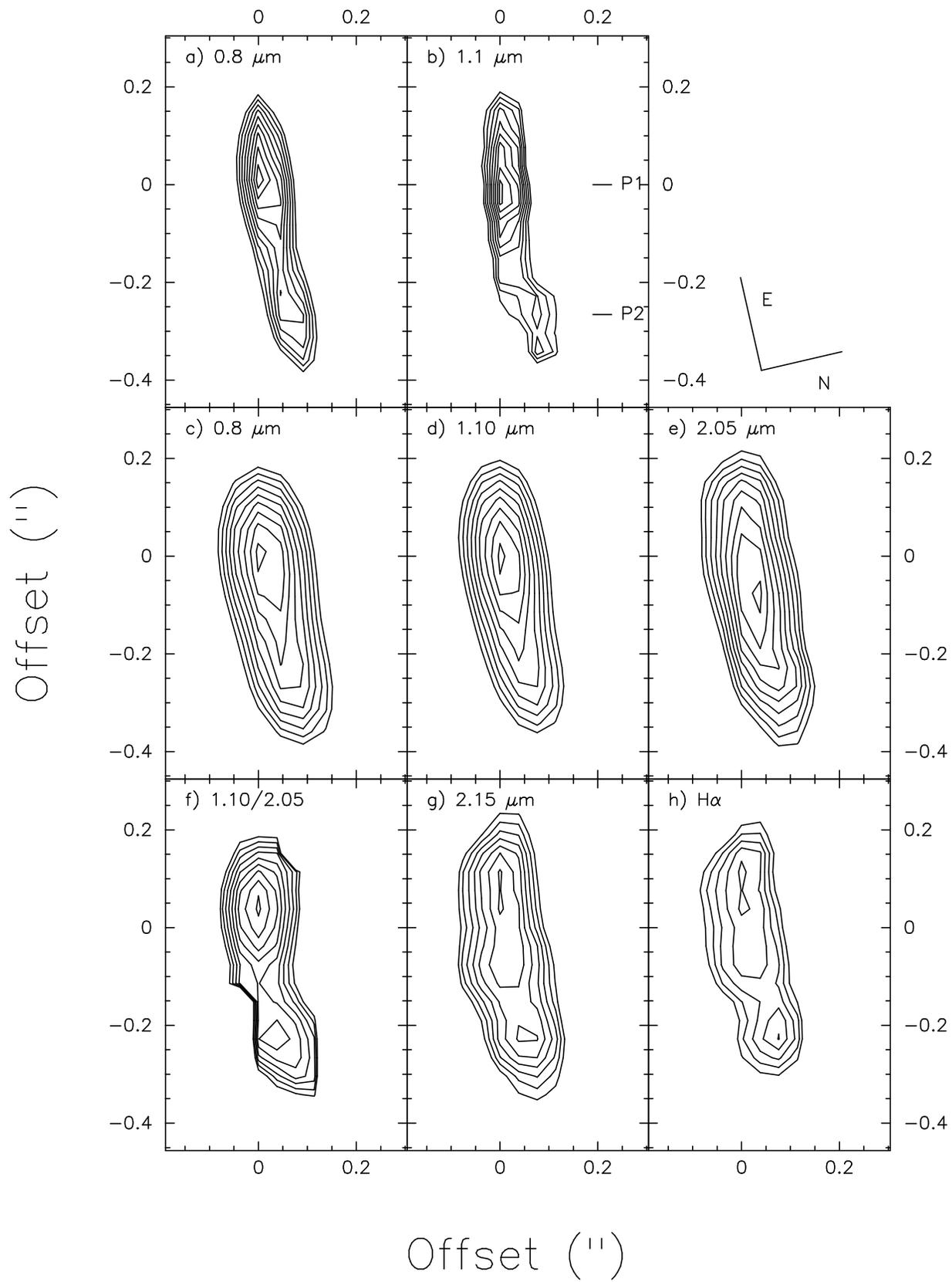
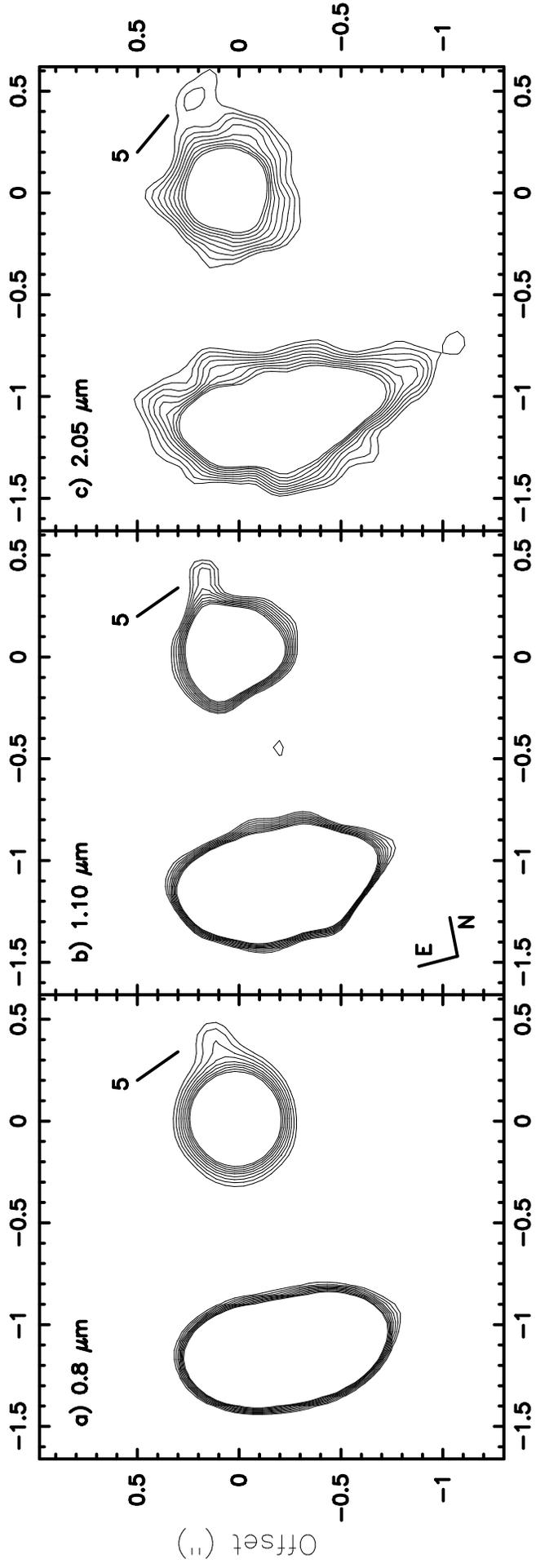
\end{document}